\documentclass[final,3p,twocolumn]{elsarticle}

\usepackage{lineno,hyperref}
	\hypersetup{colorlinks=true,pdfborder=0 0 0}
\usepackage{subcaption}
\usepackage{amsmath}
\usepackage{amssymb}
\usepackage{gensymb}
\usepackage{cleveref}
\usepackage{siunitx}
\usepackage[dvipsnames]{xcolor}
\usepackage{fancyhdr}
\modulolinenumbers[5]
\usepackage{graphicx}

\journal{Frontiers in Metals and Alloys}

\usepackage{etoolbox}
\makeatletter
\patchcmd{\ps@pprintTitle}% <cmd>
  {Preprint submitted}% <search>
  {Postprint of article submitted}% <replace>
  {}{}% <succes><failure>
\makeatother

\begin{document}

\begin{frontmatter}

\title{Learning from 2D: machine learning of 3D effective properties of heterogeneous materials based on 2D microstructure sections}

\address[az-mse]{Department of Materials Science and Engineering, University of Arizona, Tucson, AZ 85721, USA}
\address[az-am]{Graduate Interdisciplinary Program in Applied Mathematics, University of Arizona, Tucson, AZ 85721, USA}

\author[az-mse]{Guangyu Hu}
\author[az-mse,az-am]{Marat I. Latypov\corref{cor1}}
\cortext[cor1]{corresponding author}
\ead{latmarat@arizona.edu}

\begin{abstract}

Microstructure--property relationships are key to effective design of structural materials for advanced applications. Advances in computational methods enabled modeling microstructure-sensitive properties using 3D models (e.g., finite elements) based on microstructure representative volumes. 3D microstructure data required as input to these models are typically obtained from either 3D characterization experiments or digital reconstruction based on statistics from 2D microstructure images. In this work, we present machine learning (ML) approaches to modeling effective properties of heterogeneous materials directly from 2D microstructure sections. To this end, we consider statistical learning models based on spatial correlations and convolutional neural networks as two distinct ML strategies. In both strategies, models are trained on a dataset of synthetically generated 3D microstructures and their properties obtained from micromechanical 3D simulations. Upon training, the models predict properties from 2D microstructure sections. The advantage of the presented models is that they only need 2D sections, whose experimental acquisition is more accessible compared to 3D characterization. Furthermore, the present models do not require digital reconstruction of 3D microstructures. 

\end{abstract}

\begin{keyword}
Homogenization theories \sep Two-phase composites \sep Effective properties \sep Machine learning
\end{keyword}
\end{frontmatter}

\pagestyle{fancy}
\fancyhf{}
\fancyhead[LO]{Postprint of \href{ https://doi.org/10.3389/ftmal.2022.1100571}{Hu \& Latypov, Frontiers Metals Alloys (2022)}}

% \linenumbers
% Section 1

\noindent This is a postprint of the article: \
G. Hu \& M.I. Latypov,  \href{https://doi.org/10.3389/ftmal.2022.1100571}{Learning from 2D: Machine learning of 3D effective properties of heterogeneous materials based on 2D microstructure sections}, {\it{Frontiers in Metals and Alloys}} (2022). DOI:10.3389/ftmal.2022.1100571

\section{Introduction}

Multiphase alloys and metal-matrix composites constitute an important class of structural materials \cite{tasan2015overview,Latypov2016a,wang1998effect}. The presence of two or more phases allows unique combinations of properties inaccessible to single-phase materials. Accelerated design of multiphase materials and their process optimization could benefit from efficient computational tools that can accurately capture the relationships between mulitphase microstructure and overall engineering properties. 

In this context, a number of modeling approaches and strategies have been developed to date. Historically earliest, the Voigt and Reuss models \cite{voight1928handbook,reuss1929berechnung} estimated the overall or \emph{effective} properties of composites from the volume fractions of their constituents. These models are based on the assumptions of either uniform stress or uniform strain throughout the composite and serve as the lower and upper bounds of the effective property of interest (e.g., stiffness or strength). While straightforward, these bounds are often widely separated, especially in \emph{high-contrast} composites with a large difference in the properties between the constituents \cite{latypov2019computational}. Tighter bounds for effective elastic properties were obtained by Hashin and Shtrikman for composites with moderate elastic contrasts \cite{hashin1963variational}. Self-consistent models also offer better estimates of effective properties compared to bounds.  Most of these approaches, however, are based on volume fractions of the constituents, which limits their use for \emph{microstructure-sensitive} design of composite materials and structural components \cite{fullwood2010microstructure}. 

With the emergence of high-performance computing, computational homogenization has become a viable and powerful approach to calculating effective properties of heterogeneous materials \cite{segurado2018computational}. The finite element (FE) methods \cite{gilormini1987finite,ghosh1995multiple,Segurado2002,latypov2019computational} and fast Fourier solvers \cite{michel1999effective,lebensohn2001n,eisenlohr2013spectral,lucarini2021fft} both allow calculating effective elastic and inelastic properties with the direct account for three-dimensional (3D) microstructure. The microstructure effects are explicitly incorporated by performing calculations on 3D representative volume elements (RVEs) or microstructure volume elements (MVEs) of the multiphase microstructure of the material. The account for microstructure is associated with a significant increase in the computational cost compared to analytical calculations of bounds and faster numerical self-consistent methods discussed above. The computational cost of RVE-based methods is still prohibitively high for routine use in multiscale modeling of commercial metal forming processes. 

Surrogate or reduced-order models have been proposed as a strategy to address the trade-off between the computational cost and incorporation of 3D microstructure effects. Materials Knowledge Systems (MKS) is an example computational approach of surrogate model development \cite{gupta2015structure,Latypov2017}. In MKS, 3D microstructures are quantified using the $n$-point spatial statistics with subsequent establishment of quantitative microstructure--property relationships in the form of polynomial functions fitted to data from numerical (e.g., FE) simulations. Other forms of microstructure--property relationships besides polynomial functions are also seen in literature, including statistical learning models (e.g., Gaussian process regression \cite{marshall2021autonomous}), or neural networks, including convolutional neural networks, CNN \cite{yang2018deep,cecen2018material,mann2022development,ibragimova2022convolutional} and graph neural networks \cite{hestroffer2023graph,dai2021graph}. Owing to the computational efficiency and account for 3D microstructure, surrogate models offer a promising pathway towards practical implementation and industrial adoption of microstructure-sensitive multiscale models of full-scale metal forming processes. 

Both the computational homogenization methods and their data-driven surrogates require three-dimensional (3D) microstructure data in the form of an RVE as input for property calculations. The RVE needs to have sufficiently high resolution and sufficiently large size to capture both the 3D morphology and spatial distribution of the microstructure constituents (continuous phases, particles, or grains). Advances in experimental characterization have made it possible to obtain 3D microstructure datasets for a wide variety of materials. Experimental 3D datasets are typically obtained from serial sectioning \cite{groeber20063d,echlin2012new} or directly using non-desctructive 3D characterization techniques \cite{pokharel2015situ,shahani2020characterization}. Most of the experimental techniques that yield high-quality 3D microstrcuture data require access to high-cost and unique experimental facilities. Consequently high-quality 3D experimental data are currently scarce in the materials research community, especially in industry. An alternative to 3D characterization is digital generation of 3D microstructure RVEs or MVEs based on 2D microstructure data \cite{bostanabad2018computational}. For example, Dream.3D is an open-source software commonly used for 3D microstructure generation based on statistical distributions of sizes, shapes, and volume fractions of the microstructure constituents \cite{groeber2014dream}. Dream.3D has been widely used for building RVEs as input for microstructure-based simulations \cite{diehl2017identifying,latypov2017data}. Recently, new approaches have been emerging for 3D RVE generation from 2D microstructure data, including a 3D generation algorithm inspired by solid texture synthesis in computer graphics \cite{turner2016statistical} or transfer learning  \cite{bostanabad2020reconstruction}.

In this work, we propose a machine learning framework that relates the microstructure from 2D sections directly to effective 3D properties of heterogeneous materials and specifically two-phase composites. The framework is based on a hypothesis that 2D microstructure data encapsulates sufficient statistical information to capture the effective properties of heterogeneous media to a satisfactory extent. We explore the application of this framework to modeling effective mechanical properties of two-phase materials using two approaches: (i) statistical learning with spatial correlations as microstructure descriptors and (ii) deep learning with CNNs as microstructure feature extractors.  Section 2 describes these two approaches followed by their application on stiffness of two-phase microstructures presented and discussed in Sections 3--5.

\begin{figure*}[h!]
\begin{center}
\includegraphics[width=\textwidth]{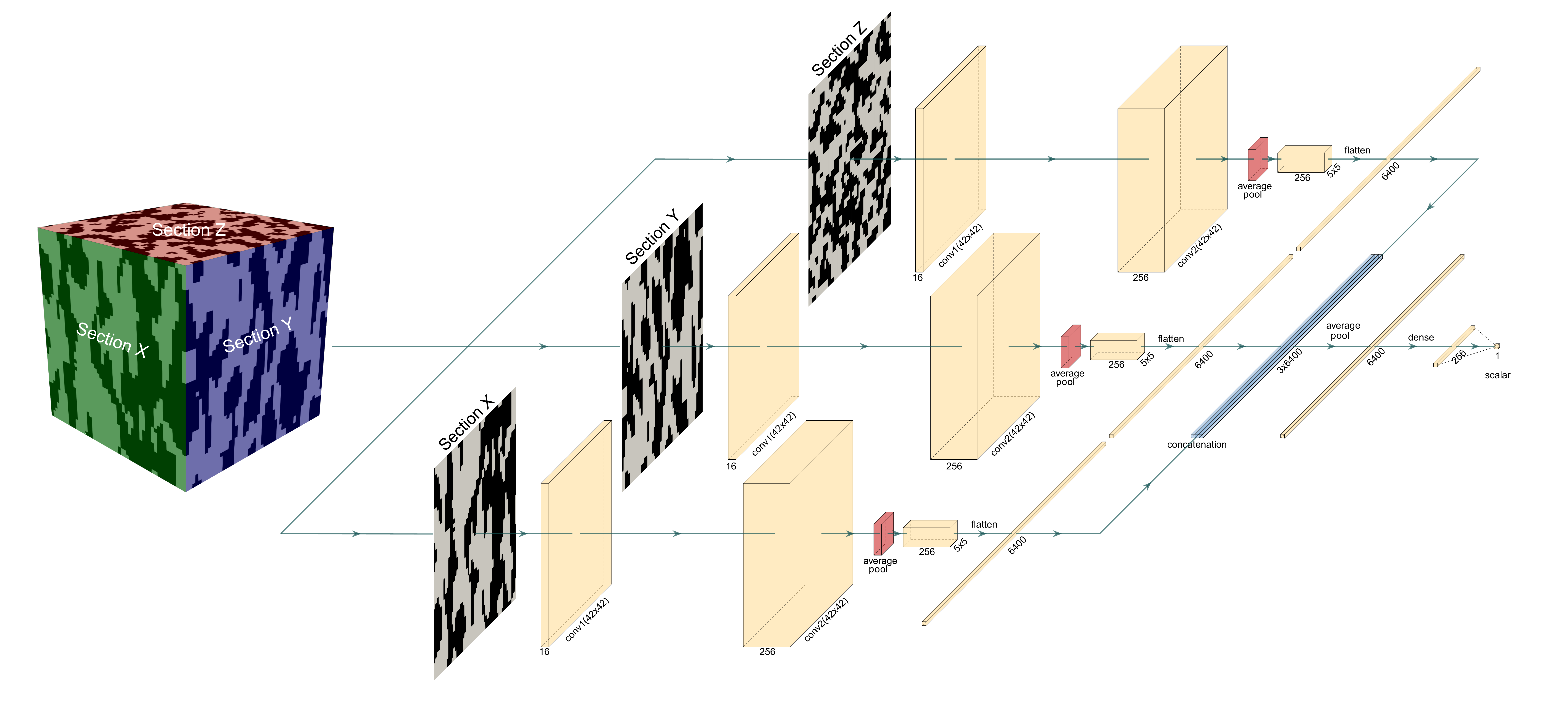}% This is a *.eps file
\end{center}
\caption{Network architecture proposed in the multisection CNN approach.}\label{fig:cnn}
\end{figure*}

\section{New Multisection modeling framework}
\label{sec:methods}

The key distinguishing feature of the computational framework developed in this study is the extraction of quantitative microstructure features from 2D orthogonal microstructure sections, their aggregation and use for machine learning models for effective properties. To this end, we develop and critically compare two approaches. In the first approach, which we refer to as the \emph{Multisection MKS approach}, we use $n$-point correlation functions and principal component analysis (PCA) to engineer microstructure features from 2D orthogonal sections. In the second, \emph{Multisection CNN approach}, we train CNNs as feature extractors for microstructure images. In both cases, features extracted from individual 2D sections are combined into a single feature vector for establishing quantitative relationships between microstructures and effective properties of interest.

\subsection{Multisection MKS approach}

As in the original developments of the MKS framework \cite{Latypov2019}, our Multisection MKS approach relies on the $n$-point correlation functions as a rigorous statistical description of heterogeneous microstructures. The theoretical basis for the use of $n$-point statistics as microstructure description in homogenization models is the classical statistical continuum theories, where an effective property is expressed as a volume-averaged quantity with corrections to higher-order microstructure effects represented by $n$-point correlation functions of increasing order with corresponding influence coefficients \cite{gupta2015structure,Brown1955,Kroner1977,Adams1989,Torquato1997}. Most of the prior MKS models were successfully developed based on two-point ($n=2$) correlation functions \cite{gupta2015structure,latypov2017data,Latypov2019,paulson2017reduced} together with PCA as a dimensionality reduction technique utilized to facilitate regression-type model calibration. In our framework, two-point correlation functions are computed individually from 2D orthogonal sections and then combined into a single vector prior to the application of PCA for dimensionality reduction. PCA is a linear transformation of multidimensional data to a new basis represented by orthogonal vectors corresponding to sequentially decreasing variance in the dataset \cite{abdi2010principal}. When PCA is applied to two-point statistics, principal component basis vectors can be interpreted as patterns in two-point correlation maps, in respect to which the spatial correlations vary the most in a set of microstructures \cite{latypov2017data,Latypov2018}. Once a new basis is established with patterns as basis vectors, the weights of these patterns are obtained as principal component scores for two-point correlations of microstructures for which prediction of the effective properties are sought. Prior MKS studies focused on 3D microstructures involved calculations of directionally-resolved 3D two-point statistics so that PCA provided 3D patterns in two-point statistics \cite{latypov2017data,Latypov2019,paulson2017reduced}. In the present computational framework, PCA yields sets of three 2D patterns in two-point statistics that correspond to the three orthogonal sections of the microstructures. Principal component scores, which represent weights of these patterns are used as features describing microstructure for machine learning of effective properties. Since PCA automatically ranks patterns according to the variance in two-point statistics, usually only a few first principal components are used for statistical learning.

\subsection{Multisection CNN approach}

Deep learning has been shown promising for predicting effective properties of heterogeneous microstructures. For modeling relationships between 3D microstructures and properties, 3D CNN architectures have been used \cite{cecen2018material}. 3D CNNs serve as extractors of microstructure features relevant for the property of interest. In the present Multisection CNN approach, we use 2D CNNs for microstructure feature extraction from three orthogonal 2D sections. The features obtained from the individual sections using separate CNNs are then combined into a single microstructure feature vector for establishing a quantitative linkage with the property of interest. Drawing inspiration from multi-view CNN methods for classification of 3D objects from 2D images taken at different angles \cite{savva2017large}, %,zeiler2014visualizing,chatfield2014return},
we employ view pooling for feature aggregation from the 2D sections. An example architecture of the Multisection CNN is shown in \Cref{fig:cnn} with two 2D convolutional layers followed by concatenation and average pooling of the features. A linear regression layer concludes the architecture for relating features from the Multisection CNN to a property of interest. While aggregation of features from individual 2D sections of the microstructure is key to our Multisection CNN approach, the other parameters of the architecture (e.g., number of convolutional layers) can be selected and tuned for each specific problem and dataset at hand. 

\section{Case study: elastic properties of two-phase microstructures}

\begin{figure*}[h]
\begin{center}
\includegraphics[width=\textwidth]{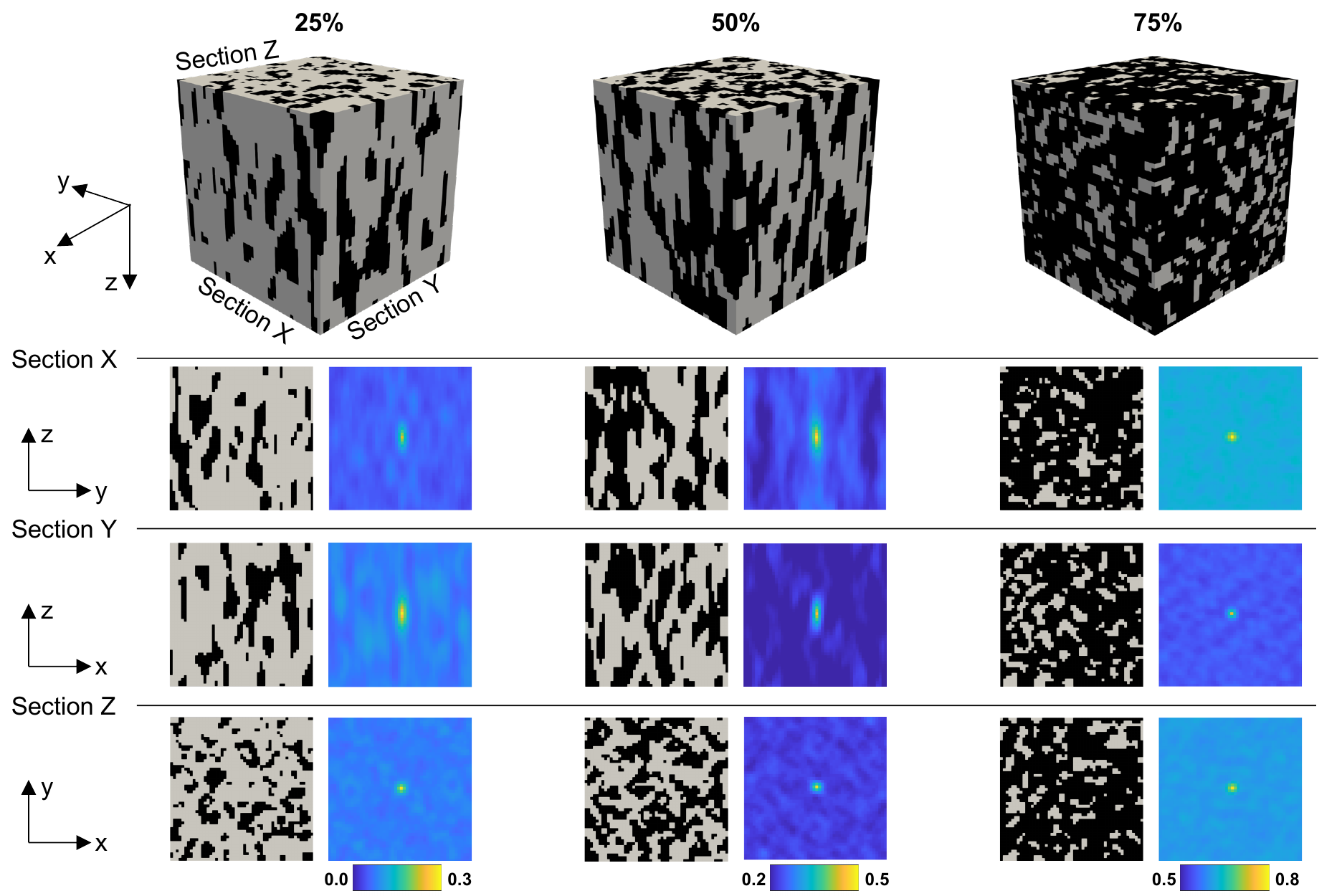}% This is a *.eps file
\end{center}
\caption{Typical 3D MVEs of two-phase microstructures contained in the dataset used for training and testing the multisection MKS and CNN models in the present study. Three orthogonal sections of the MVEs and the corresponding two-point autocorrelation maps of the stiff phase are also shown.}\label{fig:micro}
\end{figure*}

We demonstrate and evaluate the new framework in a case study of effective elastic properties in high-contrast two-phase microstructures. To this end, we leverage a dataset previously published by Cecen et al. \cite{cecen2018material}. The dataset includes 5900 two-phase microstructures and their corresponding effective Young's modulus values obtained using FE simulations. Periodic 3D MVEs of the microstructures measured $51\times51\times51$ voxels and had a volume fraction of the hard phase varied from 25\% to 75\%. The contrast of Young's modulus between hard and soft phases was 50, i.e., the hard phase was 50 times stiffer than the soft phase. \Cref{fig:micro} shows microstructures typical of the dataset. Using this dataset, we trained two new Multisection models (MKS and CNN) as well as a 3D MKS model used as a benchmark for evaluation of the new Multisection models. For all three models, the dataset of 5900 MVEs and their properties was split into a training subset (90\%) and a testing subset. The testing subset was held out during fitting of the MKS models and training of the CNN model. 

\subsection{Multisection MKS} 

The application of the Multisection MKS approach in this case study included computation of two-point correlation functions on 2D microstructure sections and PCA of the two-point statistics vectors aggregated from the three 2D sections. For two-phase microstructures, a single two-point correlation function is sufficient to fully describe the microstructure \cite{fullwood2008microstructure} so that we only calculated autocorrelation for the stiff phase using  an efficient fast Fourier transform-based algorithm \cite{cecen2016versatile}. \Cref{fig:micro} shows two-point correlation maps for 2D sections of MVEs representative of the dataset. The three concatenated 2D autocorrelations for each MVE results in a high dimensional feature vector with $3\times50\times50=7500$ elements. To facilitate machine learning, we reduced the dimensionality of the aggregated autocorrelation vector using the PCA. The selected principal component scores were then used as the final features for building a statistical learning model. In this case study, we included the first 13 principal components in a third-order polynomial function because this combination of principal components and polynomial order showed good performance for modeling effective stiffness in high-contrast two-phase microstructures \cite{cecen2018material}. 

\subsection{Multisection CNN} 

\begin{figure}[h]
\begin{center}
\includegraphics[width=0.5\textwidth]{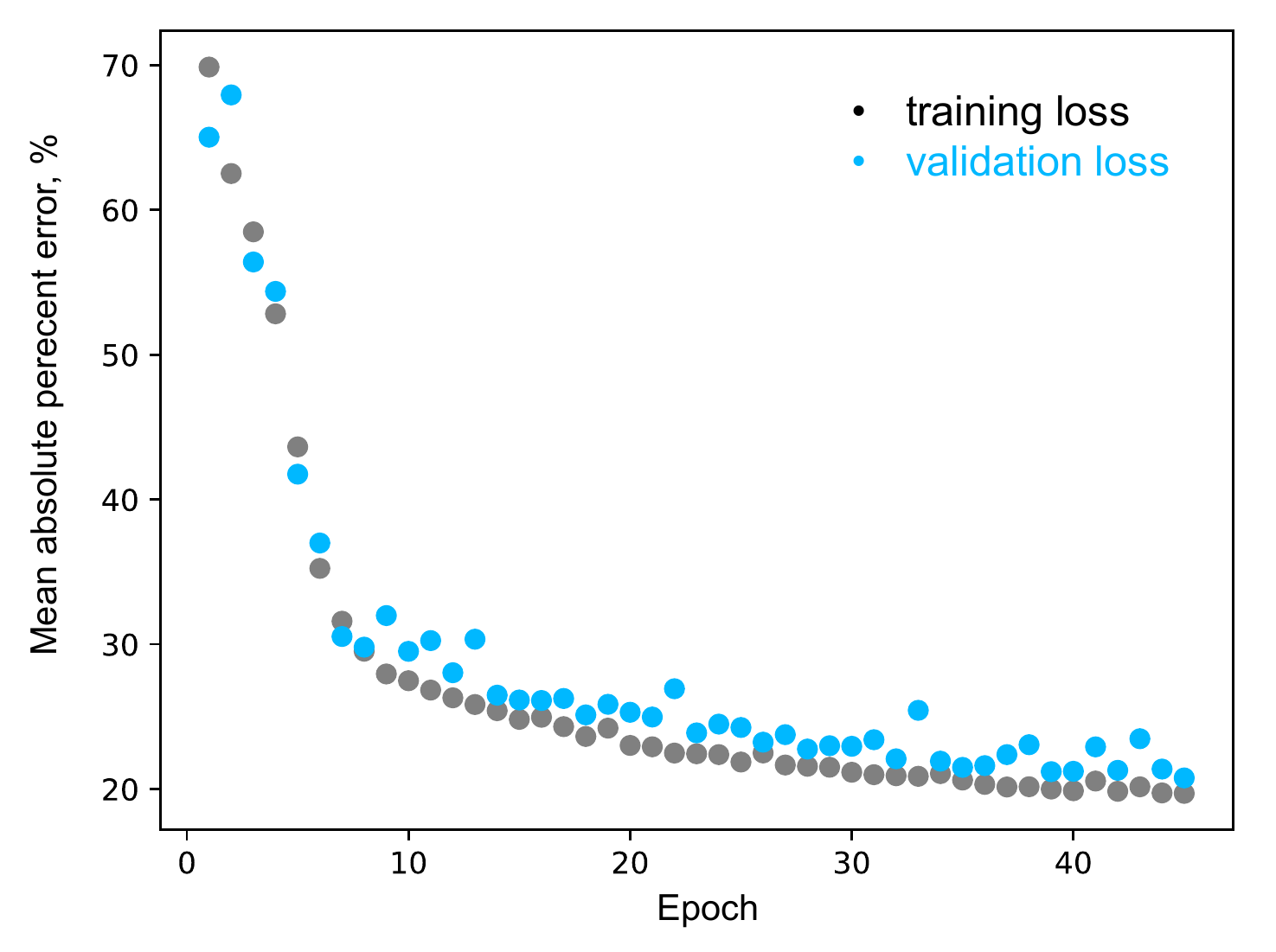}% This is a *.eps file
\end{center}
\caption{Learning curves corresponding to training of the Multisection CNN model  adopted in this study in terms of mean absolute percentage error for training and validation sets.}\label{fig:loss}
\end{figure}

Our Multisection CNN model relies on CNNs for extracting microstructure features from 2D sections. Our network architecture includes three parallel branches that take 2D image as input and then pass it through two 2D convolutional layers followed by a rectified linear layer and average pooling layer. The features produced by these three parallel channels are then further averaged to generate a single 6400-dimensional vector of features representing three orthogonal sections of the microsructure. We obtained this specific architecture by experimentation with different network configurations and by hyperparameter optimization. Bayesian optimization (on SigOpt platform \cite{dewancker2016bayesian}) was used to identify the optimal number of convolutional layers, the number of filters in the convolutional layer(s), the kernel size of the convolutional layer(s), the pooling size, batch size, learning rate, number of  for the final dense layer, the final pooling method (mean or max), optimizer (Adam or stochastic gradient descent). \Cref{fig:cnn} shows the architecture obtained by hyperparameter optimization, which includes two convolutional layers with 16 filters of $10\times10$ kernel size in the first layer and 256 filters with $4\times4$ kernel size in the second layer, followed by an average pooling layer of $8\times8$ size. We employed mean absolute error (MAE) as the loss function during training with a learning rate of $10^{-4}$. Early stopping was introduced and controlled by decrease of less than \SI{1}{\percent} in the MAE on portion of the training data. The learning curves for the model is shown in \Cref{fig:loss}. The Multisection CNN models were trained and tuned on the Ocelote cluster of the UArizona High Performance Computing system with the following specifications: two Xeon E5 14-core processors (28 CPUs total) with \SI{2.3}{GHz} and \SI{6}{GB} RAM per CPU.

\subsection{Benchmark 3D MKS} 

We additionally built a 3D MKS model which served as a benchmark for critical evaluation of the new Multisection models. We developed the benchmark 3D MKS model using a procedure similar to the Multisection MKS model. The key difference of the 3D MKS model from the Multisection MKS was that we calculated full 3D two-point correlation functions for entire volumes of the 3D MVEs. In this case, we obtained the microstructure features for machine learning by PCA of 3D two-point correlations vectors. For the benchmark 3D MKS model, we also fitted a third-order polynomial including 13 principal components of the 3D two-point correlations. 

\section{Results} 

\begin{figure*}[h]
\begin{center}
\includegraphics[width=\textwidth]{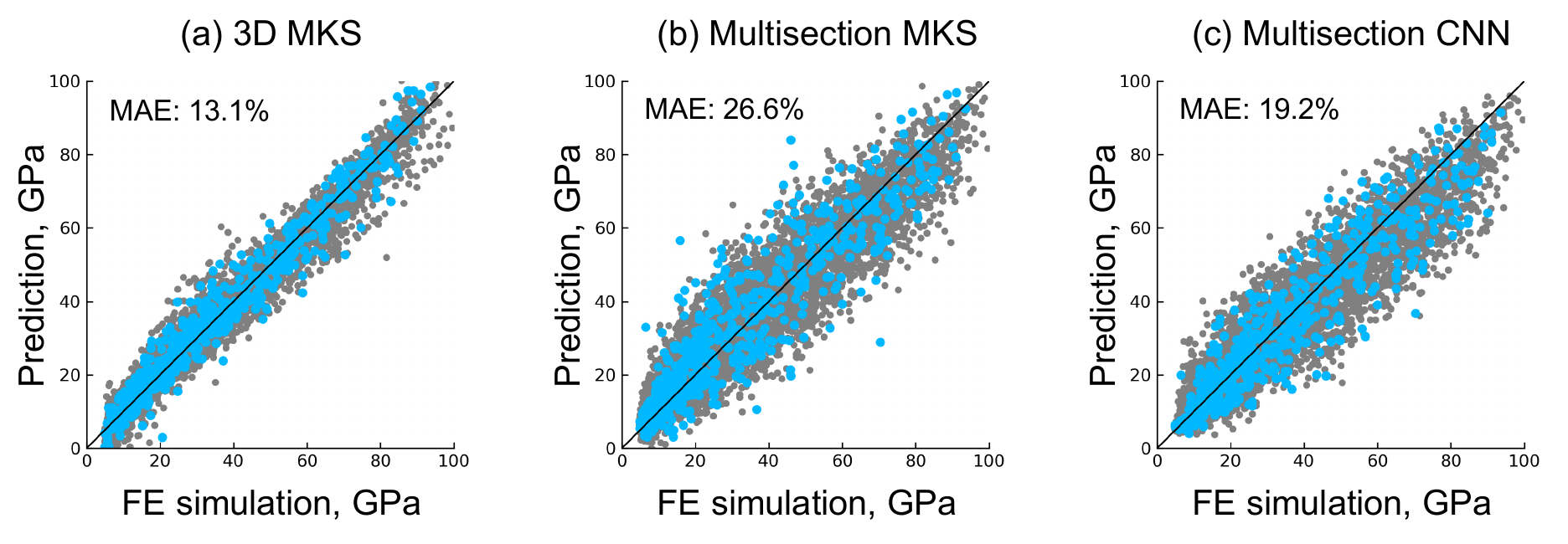}% This is a *.eps file
\end{center}
\caption{Parity plots showing the fitting results (for training set, dark grey) and the prediction results (for testing set, blue) for (a) benchmark 3D MKS model based on 3D microstructures; (b) Multisection MKS and (c) Multisection CNN models based on three 2D microstructure sections. Mean absolute errors (MAE) for the testing set normalized by ground trugh values are also shown.}\label{fig:results}
\end{figure*}

The key results of the case study with the two new models -- Multisection MKS and Multisection CNN -- are shown in the form of parity plots in \Cref{fig:results}b--c. The performance of the two models is compared to the benchmark 3D MKS model (\Cref{fig:results}a). The parity plot depicts predictions of the developed machine learning models against the ground-truth FE simulation results. Each point represents a prediction and ground truth result for one MVE; a perfect agreement would be seen as points lying on the parity line. All the models produce a reasonable agreement between the trained machine learning models and the FE simulations. Indeed, the predictions of all three models show a reasonable homoscedastic distribution along the parity line. The Multisection MKS model predictions are characterized by a wider scatter compared to the 3D MKS model, which is the ``price'' of relying on 2D microstructure information rather than full 3D data. The Multisection CNN model also shows a wider scatter compared to the benchmark model but to a smaller extent than the Multisection MKS model. The quantitative comparison of the MAE normalized by ground truth values supports these observations: the Multisection MKS model has the highest MAE of \SI{26.6}{\percent} vs.\ \SI{13.1}{\percent} of the benchmark 3D MKS model, while the MAE of the Multisection CNN is \SI{19.2}{\percent} (all for the testing set).  

\section{Discussion}

The results of the case study presented above show the feasibility of machine learning models that relate effective properties of two-phase microstructures directly to 2D microstructure sections. The value of the developed Multisection models arises from the prevalence of 2D microstructure data in the materials community, especially in industry, compared to scarce 3D data. The transition from full 3D microstructure data to 2D sections used for the MKS model development comes at the price of about \SI{13}{\percent} loss in accuracy. The trade-off between the completeness of the microstructure information and the accuracy can be addressed by more advanced models such as CNNs: we  report only a \SI{6}{\percent} increase in MAE with the Multisection CNN compared to the MKS model based on full 3D microstructure information. This sacrifice in accuracy can be reasonable in many use case scenarios given the significant savings in time, labor, and resources needed to collect 2D section data compared to 3D characterization. An additional advantage of the best performing Multisection CNN is manifested in the computational cost of training: it takes only about \SI{12}{\min} to train or Multisection CNN with CPUs (hardware specified in Section 3.1.2), while previous studies reported hours of training for 3D CNNs on the same dataset of high-contrast composites \cite{cecen2018material}. The Multisection CNN model developed here is thus a light-weight data-driven model which strikes a balance between minimal requirements in terms of microstructure input (2D sections), decent accuracy, and modest computational resources and time needed for training. 

The Multisection models presented here show a viable strategy of modeling effective properties of heterogeneous materials without the need of experimental 3D microstructure data or an extra computational step of reconstructing 3D micrsotructures from 2D data. The models trained on 2D microstructure sections from 3D synthetic microstructures used in the simulations can be leveraged for predicting properties of materials based on experimental microstructure images. While we demonstrated the predictive capabilities of the Multisection framework on effective elastic properties, it can be readily extended to effective inelastic properties of multiphase materials \cite{Latypov2017} as well as elastic and inelastic properties of polycrystalline materials \cite{paulson2017reduced}. Indeed, CNNs can be configured to take crystal orientations as input (e.g., \cite{pandey2021machine}), while calculation of two-point correlations for polycrystalline MVEs within the MKS framework is also available using generalized spherical harmonics \cite{paulson2017reduced} as the microstructure description basis. Finally, in this study, we trained a relatively simple architecture from scratch to test the concept, however, use and fine-tuning of existing state-of-the-art models (such as ResNet) as the base CNN architecture for each 2D section can futher improve the results and accuracy of the Multisection CNN approach.

\section{Summary}

We presented a new Multisection machine learning framework for predicting effective properties of heterogeneous materials. The framework is based on the hypothesis that the microstructure data from 2D sections suffice for establishing a quantitative relationship between the microstructure and the effective property of interest. We tested the hypothesis in a case study of predicting effective stiffness in high-contrast two-phase composite microstructures. To this end, we tested two computational approaches: Mutisection MKS and Multisection CNN. Both approaches rely on aggregating microstructure features extracted from three 2D sections. In the case of Multisection MKS, we obtain microstructure features by principal component analysis of two-point correlation functions calculated for 2D sections. In the case of Multisection CNN, we utilize CNN filters as feature extractors with subsequent pooling of features from individual 2D sections. We show that both the Multisection MKS and Multisection CNN models show reasonable accuracy with \SI{13}{\percent} and \SI{6}{\percent} increase in the mean average error compared to the previously published MKS model relying on full 3D microstructure data. Our results confirm the hypothesis and demonstrate the Multisection CNN approach as an optimal model for predicting effective properties of heterogeneous media without the need to collect or reconstruct 3D microstructure volumes. The Multisection CNN is therefore a powerful addition to the computational materials science toolkit for modeling microstructure--property relationships.

\section*{Acknowledgments}

Our machine learning models were trained and tuned using High Performance Computing (HPC) resources supported by the University of Arizona TRIF, UITS, and the Office of Research, Innovation, and Impact, and maintained by the UArizona Research Technologies Department.

\bibliography{refs}

\end{document}